# MAPPING OF FOCUSED LAGUERRE-GAUSS BEAMS: THE INTERPLAY BETWEEN SPIN AND ORBITAL ANGULAR MOMENTUM AND ITS DEPENDENCE ON DETECTOR CHARACTERISTICS

**V.V. Klimov**

*P.N. Lebedev Physical Institute, Russian Academy of Sciences,*

*53 Leninsky Prospekt, Moscow 119991, Russia*

*e-mail: vklim@sci.lebedev.ru*

**D. Bloch, M. Ducloy**

*Laboratoire de Physique des Lasers, Université Paris 13, Sorbonne Paris-Cité*

*and*

*CNRS, UMR 7538, 99 Ave J.B. Clément, F-93430 Villetaneuse, France*

*e-mail : daniel.bloch@univ-paris13.fr*

**J. R. Rios Leite**

*Departamento de Física, Universidade Federal de Pernambuco,*

*50670-901 Recife, PE, Brazil*

**Abstract**

*We show that propagating optical fields bearing an axial symmetry are not truly hollow in spite of a null electric field on-axis. The result, obtained by general arguments based upon the vectorial nature of electromagnetic fields, is of particular significance in the situation of an extreme focusing, when the paraxial approximation no longer holds. The rapid spatial variations of fields with a "complicated" spatial structure are extensively analyzed in the general case and for a Laguerre-Gauss beam*

*as well, notably for beams bearing a |l| = 2 orbital angular momentum for which a magnetic field and a gradient of the electric field are present on-axis. We thus analyze the behavior of a atomic size light-detector, sensitive as well to quadrupole electric transitions and to magnetic dipole transitions, and apply it to the case of Laguerre-Gauss beam. We detail how the mapping of such a beam depends on the nature and on the specific orientation of the detector. We show also that the interplay of mixing of polarization and topological charge, respectively associated to spin and orbital momentum when the paraxial approximation holds, modifies the apparent size of the beam in the focal plane. This even leads to a breaking of the cylindrical symmetry in the case of a linearly polarized transverse electric field.*

# 1. Introduction

The angular momentum of light has been a subject of active research for many decades and it is frequently revived by novel theoretical and experimental techniques in optics. The light beams with orbital angular momentum introduced nearly twenty yeas ago [1,2] (for reviews, see [3]) in the frame of a classical radiation interpretation, have triggered one of these intense interest periods. The same period has seen the advent of well controlled sources of photons also carrying orbital angular momentum. Despite so much work on the subject, many issues are still under debate as to the interpretation of the orbital angular momentum of light as classical radiation and photons [4,5]. The subject has direct implication on the manipulation of matter by light and quantum information applications, where orbital momentum may provide an extra-degree of entanglement [6] The qubits associated to photon intrinsic polarization, spanning a two dimensional Hilbert space, seem to be of a quite different nature as compared to those based upon the quantum states of the orbital angular momentum quantum space (see for instance [7]). The difference is inherited from classical mechanics where spin, *i.e.* intrinsic angular momentum of a rigid body, is independent of external axis orientation. Conversely, the orbital angular momentum of any particle does depend on the reference space axis orientation.

For light radiation, an orbital angular momentum is defined when a fixed axis is defined as the axis of the light beam. The plane wave or paraxial solutions of Maxwell's equations can have its quantized version with photons expanded in a base of vectorial Harmonics [8], whose quantum numbers indicates spin and orbital angular momentum. When the solution of classical Maxwell equations is obtained in the paraxial approximation and expanded onto Laguerre-Gauss (LG) functions, there is no need to transform plane wave into spherical harmonics. The solution brings directly the integer number that indicates the orbital twist of the beam, which is associated to the orbital angular momentum along the beam axis [1]. This provides, for paraxial light beams, a simple interpretation for photons as quanta with spin and orbital angular momentum: the total angular momentum is just the component along the propagation axis. The non-paraxial solutions with Bessel functions are not so simple to view as composed of photons with specific spin and orbital quantum numbers. The spatial axis corresponding to the propagation direction cannot be approximated as unique in a wave front, and the interpretation of the quanta as having a projected orbital component along the central axis direction is not free of controversy. It was in this spirit that we have started to investigate in detail the interaction of a LG beam with multipolar atomic transitions, as a probe of the elementary transfer (at the photon level) of energy and angular momentum, and as a way to perform all the transitions allowed by selection rules [7,9,10].

In this initial work [9], we had shown that in Laguerre-Gauss (LG) beams, energy, and hence photons, can be in principle detected in regions where there is no electric field. In such a situation, a standard point-like electric dipole (E1) detector is blind. Our demonstration was based upon a development on a Bessel beam expansion [9], and on the fact that a null on-axis electric field does not imply a zero magnetic field, nor zero higher-order field gradients, on this symmetry axis. On these grounds, we had emphasized the interest to implement magnetic detectors at the optical frequency (for a recent experimental demonstration, see [11]), or "gradient" optical detectors, based upon high order transitions of a quantum atomic system although these transitions are often considered to be nearly forbidden in the optical frequency domain.

Extending our previous results, the present paper analyzes, in a general formalism, the spatial structure of "complicated" electromagnetic fields –*i.e.* fields exhibiting rapid spatial variations on a wavelength scale-, notably those with an axial symmetry. This is applied in detail to LG beams, allowing us to visualize explicitly the (vectorial) electric and magnetic fields. A reason to go to general methods is that LG fields are only an approximate solution of the Maxwell equations, valid in the framework of the paraxial approximation. Indeed, the complicated spatial structure of optical beams, including LG beams, is mostly apparent in the conditions of strong or ultimate focusing, at odds with the conditions allowing the paraxial approximation. To predict the specific features of a field bearing a complicated structure, one has to evaluate the interaction of such a field with a specific high-order quantum detector (*i.e.* a detection not based upon an E1 transition). For this purpose, we establish on a general basis the excitation rate of an elementary atomic system by an optical field, assuming that the interaction can be restricted to the electric quadrupole (E2) and magnetic dipole (M1) transitions, in addition to the standard electric dipole (E1) transitions. We hence derive, at the center of a LG beam, general formulae for the interaction with a high-order detector, allowing the selective excitation of given components of the detecting transition: the response of the detector (whose nature is tensorial), depends upon its orientation and polarization of the complicated field, and appears sharply enhanced with focusing. We also provide a detailed mapping of a focused LG beam as a function of the type of detector, showing that, because of the longitudinal components of the field, the detector-dependent mapping is highly sensitive to the combined choice of polarization and topological charge. This combination is susceptible to govern the apparent size of the beams, eventually leading to a break of the cylindrical symmetry for a linear polarization .

The paper is organized as follows: (i) in section 2, after a brief discussion of the usual definition of "hollow beam" (sub-section 2.1), we detail our calculation of the field structure in a Bessel beam, known to be an exact solution of Maxwell equations (sub-section 2.2). This calculation includes the specific evaluation of the electric and magnetic fields, found to be nonzero on-axis as long as the angular momentum is equal to 2. We extend this calculation to LG beams (sub-section 2.3). In Section 3, we first present (sub-section 3.1) the theory of the excitation of an atomic quantum system by an optical field with a complicated spatial structure, allowing a simultaneous excitation on a E1, M1, and E2 transitions, and we apply it to the case of a LG beam in sub-section 3.2.

## 2. Spatial properties of the electric and magnetic fields for axially symmetric light beams

### 2.1 Axial symmetry and vectorial "hollow" beams

Usually, it is thought that the hollow structure of a beam originates in an axial symmetry argument. Indeed, any scalar field with phase factor $e^{il\varphi}$ (with $l$ a nonzero integer) should have a null amplitude on the axis of symmetry to provide a unique value of the field. However electromagnetism deals with vectorial fields **E** and **B** and, in the general case, there is no solution of Maxwell equations, where all the components of **E** and **B** are proportional to the same factor $e^{il\varphi}$.

Indeed, if we consider a monochromatic **E** field at circular frequency $\omega$ with the following structure:

$$\mathbf{E} = \{E_x(r,z), E_y(r,z), E_z(r,z)\} e^{il\varphi} \qquad (1)$$

where $r = \sqrt{x^2 + y^2}, \varphi = \arctan(y/x)$, we find for the magnetic field $ik\mathbf{B} = rot\mathbf{E}$ with $k=\omega/c$:

$$B_x = -\frac{i}{k}\left(\frac{\partial E_z(r,z)}{\partial r}\sin\varphi + \frac{E_z(r,z)}{r}il\cos\varphi - \frac{\partial E_y(r,z)}{\partial z}\right)e^{il\varphi} \neq B_x(r,z)e^{il\varphi} \qquad (2)$$

In the absence of charges, *i.e.* when div**E**=0, one can demonstrate that only 2 components of (1) have the same axial symmetry. So, one can at most assume that only the transversal components of **E** (or **B**) field have the same symmetry, and the electric field can be represented in the form

$$\mathbf{E} = \{E_x(r,z), E_y(r,z), E_z(r,z,\varphi)\} e^{il\varphi} \qquad (3)$$

where $E_z(r,z,\varphi)$ has a nontrivial dependence on $\varphi$.

This reasoning proves that vectorial Maxwell equations do not allow a solution with a full axial symmetry. As a result, Maxwell's equations admit that some components of **E**, or **B** = - i/k rot**E**, or higher gradients of fields can be nonzero on the symmetry axis of a propagating field. This fact explains why there is no truly hollow beam in electromagnetism [12]. More specifically, when LG and Bessel beams are considered to be hollow beams, this actually applies solely to the electric field, not necessary to the magnetic field, higher-order gradients, or e.m. field intensity.

## 2.2. Near axis properties of generalized vectorial Bessel beams

For a direct demonstration of the nontrivial on-axis structure of freely propagating beams, let us start from an exact solution of Maxwell equation in free space for which we assume a cylindrical symmetry through an $e^{il\varphi}$ phase factor, *i.e.* from the so-called generalized Bessel beam [2,9].

The Cartesian components of the electric field are represented in cylindrical coordinates (*r, φ, z*) as:

$$\{E_x, E_y\} = \{\alpha, \beta\} e^{il\varphi} \int_0^k d\kappa\, g(\kappa) e^{ihz} J_l(\kappa r)$$

$$E_z = \frac{1}{2} e^{il\varphi} \int_0^k d\kappa\, g(\kappa) \frac{\kappa}{h} e^{ihz} \begin{bmatrix} (i\alpha - \beta) e^{-i\varphi} J_{l-1}(\kappa r) \\ -(i\alpha + \beta) e^{i\varphi} J_{l+1}(\kappa r) \end{bmatrix}$$

(4)

where $J_l$ is the Bessel function of order $l$, $h = \sqrt{k^2 - \kappa^2}$ is the longitudinal wavenumber, $k=\omega/c$ is the free space wave number with ω the light (circular) frequency. In (4), $g(\kappa)$ is an arbitrary function, and $\alpha$ and $\beta$ are the (electric field) polarization components (assumed to be constant in a transverse plane, with $\alpha^2 + \beta^2 = 1$).

Through the choice of $g(\kappa)$, a wide range of desirable distributions in $r$ and $z$ can be considered. For example, for

$$g(\kappa) = \exp\left(-\frac{\kappa^2 z_R}{2k}\right)\left(\frac{\kappa}{k}\right)^{(2p+|l|+1)} \tag{5}$$

eq. (4) describes the so called "elegant" Laguerre-Gauss beams in the limit of large Rayleigh numbers $z_R$ [2]. In (4), the integration is limited to propagating waves, $\kappa < k = \omega/c$. However, if generalized Bessel beams are produced in a nanoenvironment, one should also take into account evanescent waves with $\kappa > k = \omega/c$. Equations (4) strictly satisfy the transversality condition div**E** = 0. The components of the magnetic field are derived from the Faraday law as :

$$B_x = \frac{c}{2\omega}\int_0^k d\kappa \frac{g(\kappa)}{2h} e^{ihz+il\varphi} \left\{ \begin{array}{l} \kappa^2\left((i\alpha-\beta)e^{-i2\varphi}J_{l-2}(\kappa r)-(i\alpha+\beta)e^{i2\varphi}J_{l+2}(\kappa r)\right) \\ -2\beta(2h^2+\kappa^2)J_l(\kappa r) \end{array} \right\}$$

$$B_y = -\frac{c}{2\omega}\int_0^k d\kappa \frac{g(\kappa)}{h} e^{ihz+il\varphi} \left\{ \begin{array}{l} \kappa^2\left((\alpha-i\beta)e^{i2\varphi}J_{l+2}(\kappa r)+(\alpha+i\beta)e^{-i2\varphi}J_{l-2}(\kappa r)\right) \\ +2\alpha(2h^2+\kappa^2)J_l(\kappa r) \end{array} \right\}$$

$$B_z = -\frac{ic}{2\omega}\int_0^k d\kappa\, g(\kappa)\kappa e^{ihz+il\varphi}\left[(-i\alpha+\beta)e^{-i\varphi}J_{l-1}(\kappa r)-(i\alpha+\beta)e^{i\varphi}J_{l+1}(\kappa r)\right]$$

(6)

According to [2] energy of Bessel beam per unit length is

$$\mathcal{E} = \frac{1}{8}\int_0^k d\kappa \frac{|g(\kappa)|^2 (2k^2 - \kappa^2)}{\kappa(k^2 - \kappa^2)} \tag{6}$$

while z- component of the total angular momentum per unit length is

$$\mathcal{L}_z = \left[l + i(\alpha^*\beta - \alpha\beta^*)\right]\frac{1}{8\omega}\int_0^k d\kappa \frac{|g(\kappa)|^2 (2k^2 - \kappa^2)}{\kappa(k^2 - \kappa^2)}$$

$$+ i(\alpha^*\beta - \alpha\beta^*)\frac{1}{8\omega}\int_0^k d\kappa \frac{|g(\kappa)|^2 \kappa}{(k^2 - \kappa^2)} \tag{6}$$

From the ratio between Eqs. (6) and (6) one obtains

$$J_z = l + i(\alpha^*\beta - \alpha\beta^*) + \Delta J_z$$

$$\Delta J_z = i(\alpha^*\beta - \alpha\beta^*)F < 1 \tag{6}$$

$$F = \frac{\int_0^k d\kappa \frac{|g(\kappa)|^2 \kappa}{(k^2 - \kappa^2)}}{\int_0^k d\kappa \frac{|g(\kappa)|^2 (2k^2 - \kappa^2)}{\kappa(k^2 - \kappa^2)}} < 1$$

It follows from (6) that in arbitrary case of Bessel beam it describes the beam with angular momentum, which is only approximately equal to $l + i(\alpha^*\beta - \alpha\beta^*) = l + \sigma_z$. The accuracy of this approximation becomes better for large values of orbital momentum ($l \gg 1$) and/or for weaker focusing ($\frac{1}{kw_0} \to 0$). In our case, and in spite of the relatively strong focusing of the beam ($kw_0 = 6$), one has however $(kw_0)^{-1} \ll 1$ and $l=2 >1$, so that one can expect that the Bessel beam has only one component with a well-defined orbital number.

Now let us investigate the asymptotic behavior of all field components near axis ($r \to 0$, i.e. $r \ll \lambda$) in the general case (a brief description for $|l| = 0,1,2$ was already given in [9]). These asymptotes can be easily found from (4)-(6) by making use of the expansion of Bessel functions for small arguments: $J_l(x) \approx \left(\frac{x}{2}\right)^l (1 + O(x^2))$

For $l=0$, whatever is the choice of $g(\kappa)$, the non zero fields are:

$$E_x(r=0) = \alpha \int_0^k d\kappa\, g(\kappa) e^{ihz}$$

$$E_y(r=0) = \beta \int_0^k d\kappa\, g(\kappa) e^{ihz}$$

$$B_x(r=0) = -\frac{1}{2}\int_0^k d\kappa\, g(\kappa)\frac{\beta}{hk}\left(2h^2+\kappa^2\right)e^{ihz}$$

$$B_y(r=0) = \frac{1}{2}\int_0^k d\kappa\, g(\kappa)\frac{\alpha}{hk}\left(2h^2+\kappa^2\right)e^{ihz}$$

(7)

so that the field just exhibits a quasi plane-wave transversal structure near axis.

For $|l|=1$ the only possible nonzero components are:

$$E_z(r=0) = \frac{1}{2}\int_0^k d\kappa\, g(\kappa)\frac{\kappa}{h}(-\beta\pm i\alpha)e^{ihz}$$

$$B_z(r=0) = -\frac{1}{2}\int_0^k d\kappa\, g(\kappa)\frac{\kappa}{h}(\alpha\pm i\beta)e^{ihz} \quad (7)$$

In (7), $-\beta\pm i\alpha$, and $\alpha\pm i\beta$, stands for $l=\pm 1$. Hence, only a longitudinal field exists on axis, resulting from the wavefront curvature. Note that these longitudinal contributions retain a dependence on the polarization coefficients $\alpha$ and $\beta$ (they can be accidentally null). The $E_z$ and $B_z$ terms vanish for a quasi-plane wave structure, when $g(\kappa)$ is nonzero only for small transversal wavenumbers [$g(\kappa)\approx\delta(\kappa)$; with $\delta$ the Dirac function].

For $|l|=2$, only the transversal components of B are nonzero on-axis :

$$B_x(r=0) = \pm\frac{1}{2}\int_0^k d\kappa\, g(\kappa)\frac{i\kappa^2}{2hk}\left[(\alpha \pm i\beta)\right]e^{ihz}$$

$$B_y(r=0) = \pm i B_x(r=0)$$

(7)

In (7), $(\alpha \pm i\beta)$ stands for $l = \pm 2$.

For $|l| > 2$, all components of **E** and **B** are zero on-axis, but higher-order gradients of fields can be nonzero even in this case.

Hence, the generalized Bessel beams have nonzero longitudinal fields $E_z, B_z$ on-axis for $|l|=1$. For $|l|=2$, they have an on-axis magnetic field, which exhibits a transverse structure and a circular polarization independent of the polarization components $\alpha$, $\beta$, *i.e.* the sign of this circular polarization depends only on the sign of orbital momentum $l$ of optical beam. Note that it is only for one of the circular polarization of the transverse electric field that this on-axis magnetic field for $|l|=2$ turns to be null (*e.g.* for $\alpha + i\beta = 0$ when $l = +2$). It also yields a non-zero magnetic energy density on-axis:

$$I_M = \frac{|B_x|^2 + |B_y|^2}{16\pi} = \frac{1}{128\pi}\left|\int_0^k d\kappa\, g(\kappa)\frac{\kappa^2}{hk}\right|^2 |\alpha \pm i\beta|^2$$

(7)

in spite of an electric energy density equal to zero on–axis. This makes this region essentially different from any local region of traveling plane waves, which have equal amounts of energy density. Although often unnoticed, this situation is actually very common, as exemplified with the nodes of a standing wave (see [11]), making standing waves a kind of "complicated" field (*i.e.* with a subwavelength structure).

This extension of the demonstration in [9] confirms our general statement that "hollow" beams with $|l|=2$ are not truly hollow.

In a similar manner, but with a higher complexity as due to its tensorial nature, one can demonstrate that, for $|l|=2$, the gradient of the electric field is non zero on the axis (and so on, for arbitrary $l$, if one considers successive derivations of the field at the adequate order).

With this example of generalized Bessel beams, we have shown that axially symmetric beams have nonzero field components on axis. This result is actually not restricted to Bessel beams. Any solution of Maxwell equation with radial symmetry of transversal electric (or magnetic) fields, necessarily have nonzero field components or derivatives on the axis. Below, we develop this point for the case of LG beams.

## 2.3. Properties of vectorial Laguerre-Gauss beams

In this subsection, we analyze specifically the situation of a Laguerre-Gauss (LG) beam, which is a "hollow" beam of a particular interest, as due to the fact that such a beam usually carries an orbital angular momentum [1,13]. The electric field of a LG beam can be described by the following formulae [13]:

$$\mathbf{E}^{(l)}(\mathbf{r},\omega) = E_0 \frac{w_0}{k} \left\{ k\alpha U^{(l)}, k\beta U^{(l)}, i\left(\alpha \frac{\partial U^{(l)}}{\partial x} + \beta \frac{\partial U^{(l)}}{\partial y}\right) \right\} e^{ikz}$$

$$U_p^{(l)} = \frac{C_p^{|l|}}{w(z)} \left[\frac{\sqrt{2}r}{w(z)}\right]^{|l|} \exp\left(-\frac{r^2}{w^2(z)}\right) L_p^{|l|}\left(\frac{2r^2}{w^2(z)}\right) \times \qquad (7)$$

$$\exp\left(\frac{ikr^2 z}{2(z^2+z_R^2)} - il\varphi - i(2p+|l|+1)\arctan(z/z_R)\right)$$

where $C_p^{|l|} = \sqrt{2p!/\pi(p+|l|)!}$ is the normalization constant, $w(z) = w_0\sqrt{1+z^2/z_R^2}$ is the beam radius at $z$, $w_0$ is the Gaussian beam waist, $L_p^{|l|}(x)$ is the generalized Laguerre polynomial of order $p$, index of angular rotation $l$, and argument $x$. The Rayleigh range of the beam is $z_R = kw_0^2/2$, $(2p+|l|+1)\arctan(z/z_R)$ is the Gouy phase and, as in (4), $(\alpha,\beta)$ are the polarization vector components with $|\alpha|^2 + |\beta|^2 = 1$. Note that $p+1$ is the number of nodes of the field in the radial direction and that $l$ is the orbital angular momentum per photon carried by the beam along its propagation direction in units of $\hbar$ [1].

We notice that the LG modes are proportional to $r^{|l|}e^{-il\varphi} = (x \pm iy)^{|l|}$, a term characteristic of the eigenfunctions of the orbital angular momentum operator $l_z = -i\hbar \qquad \hbar$. Laguerre-Gauss beams bear both spin and orbital angular momentum. Their total averaged momentum per quantum of energy $\hbar\omega$ is, by :

$$j_z = \hbar \qquad (7)$$

where $\sigma = -i(\alpha\beta^* - \beta\alpha^*)$ can be considered as a quantum expectation value of spin operator. Note that eq. (7) is only an approximate expression, as for highly focused beams, one should use a more precise expression for LG beams [2] (see also eq. (6)) : there are numerous discussions in the literature about the orbital/spin separation when an off-axis measurement is considered.

The electric field vector of eq. (7) has a longitudinal *z*-component [13]. The presence of longitudinal fields is necessary to provide the charge conservation law div**E**=0. Strictly speaking, even eq. (7) does not satisfy the charge conservation law because div**E** ~ $1/k^2 \neq 0$. The definition of LG beams can however be made more precise [14], but in what follows, and with respect to our order of approximation, there is no necessity to do it.

The magnetic field, derived from eq. (7) through the Faraday's law of induction $ik\mathbf{B} = rot\mathbf{E}$, has the following form:

$$\mathbf{B}^{(l)}(\mathbf{r},\omega) = E_0 \frac{w_0}{k} \left\{ \begin{array}{l} -k\beta U^{(l)} + i\beta \dfrac{\partial U^{(l)}}{\partial z} + \dfrac{\alpha}{k} \dfrac{\partial^2 U^{(l)}}{\partial x \partial y} + \dfrac{\beta}{k} \dfrac{\partial^2 U^{(l)}}{\partial y^2}, \\ k\alpha U^{(l)} - i\alpha \dfrac{\partial U^{(l)}}{\partial z} - \dfrac{\alpha}{k} \dfrac{\partial^2 U^{(l)}}{\partial x^2} - \dfrac{\beta}{k} \dfrac{\partial^2 U^{(l)}}{\partial x \partial y}, \\ i\left( \alpha \dfrac{\partial U^{(l)}}{\partial y} - \beta \dfrac{\partial U^{(l)}}{\partial x} \right) \end{array} \right\} e^{ikz} \quad (7)$$

As in the case of Bessel beams, there can be magnetic energy density (see eq.(7) ) and gradients of electric field on-axis [9], or a longitudinal electric field (for |*l*| = 1).

For |*l*| =2 the nonzero components are the following :

$$B_x(0) = \mp \frac{\sqrt{8(p+1)(p+2)}}{k\sqrt{\pi}w_0} \mp \quad \pm iB_x(0)$$

$$\frac{\partial E_z}{\partial x}(0) = i\frac{\sqrt{8(p+1)(p+2)}}{\sqrt{\pi}w_0^2 k}(\alpha \mp \quad \quad \mp$$

(7)

From (7) one sees again (see section 2.2) that the on-axis magnetic field always exhibits a circular polarization solely governed by the sign of *l*, even for a linear polarization of the transverse electric field. It is only if the transverse electric field is circularly polarized with the same sign as the topological charge *l* that this on-axis magnetic field turns to be null. A similar behavior appears for the gradient of the electric field. These non zero magnetic field and gradient of the electric field on-axis imply a non-zero response for an atomic detector based upon a M1 or E2 transition, while remarkably a standard photon detector, based upon the detection of an E1 transition, cannot work here. This naturally implies that in this case the electric energy density is zero on the axis, while the magnetic energy density $I_M$ is nonzero [9]:

$$I_M = \frac{1}{16\pi}|B|^2 = E_0^2 \frac{(p+1)(p+2)}{\pi^2(kw_0)^4}|\alpha \mp \qquad (7)$$

In (7), $(\alpha \mp$ stands for $l = \pm 2$.

Equation (7) shows that it is for strongly focused beams, and for high *p* values, that the magnetic energy density can be substantial. As already mentioned in [9] and as evidenced by figs. 1 and 2, for a tight focusing, the magnetic energy on-axis can compete with the electric energy density existing only off-axis. In Fig. 2, this ratio of the magnetic energy density on-axis to the energy density of the transversal electric field at its maximum is shown as a function of the beam waist (solid line), both for a LG beam as described in eq.(7), and for an exact solution with a Bessel beam [eq. (4)]. The behavior for both beams is very similar and supports the validity of our LG beam description above the paraxial approximation, even in the case of strong focusing.

Figures 3-4 show the distributions in the waist plane of the electric and magnetic fields, for the respective case of $\sigma=-1$ and $\sigma=+1$ polarizations and $l = +2$. As can be expected following eq.(7), major differences are observed between the two circular polarizations. Essentially, the predicted field structure exhibits a one-fold axial symmetry for a $\sigma=-1$ and $l=+2$ situation, *i.e.* $j_z/\hbar = +1$, while a three-fold axial symmetry appears for a $\sigma=+1$ and $l = +2$ situation, *i.e.* $j_z/\hbar = +3$ . Note that figs. 3 and 4 would be inverted for $l=-2$, to respect the symmetry imposed by the $|j_z/\hbar|$ value. If an arbitrary point in the waist plane is considered, the total electric field, and the magnetic field as well, exhibit in most cases an elliptical polarization. For some peculiar circles (*e.g.* node of the longitudinal magnetic field), one notes that the magnetic field oscillates radially, or tangentially, with however a phase that depends upon the azimuthal angle.

In figs. 5 and 6, the equivalent distributions are plotted for a linear (*x*) polarization and appear to be even more complex. Although the linear polarization is nothing else than the summing of a $\sigma=-1$ and $\sigma=+1$ polarized fields, the difference in the symmetry of the two principal circular polarizations, according to the $|j_z/\hbar|$ value, implies peculiar features: in particular, instead of a one-fold or three-fold axial symmetry along an axis that rotates at the optical frequency, all the transverse components of the field are characterized by a two-fold symmetry defined along the fixed *x, y* axes. Also, the longitudinal components of the field exhibit a complex temporal evolution, as resulting from the sum of clockwise (fig. 3), and anticlockwise (fig. 4) rotation.

The above discussion is about the field-structure in the waist plane. When moving away from the focal plane by a fraction of the Rayleigh length, more complicated structures are expected to appear, because there is no effective planar symmetry to be expected in the focal "plane" for these "spiral" fields.

## *3. Mapping complicated vectorial optical fields with elementary atomic detectors*

We have already defined (in section 1) "complicated" field as a field whose spatial variations and gradients are strong on a wavelength scale, hence beyond the plane wave approximation. Calculating the interaction of a quantum system with a "complicated" vectorial field requires a description of the interaction of that system with the different orders of the gradients of the electromagnetic field. Here, we first present (sub-section 3.1) the general expression yielding the excitation rate when the coupling between the gradients of the optical fields and the multipole momenta of the atomic or molecular system is not neglected, going beyond the usual limits of electric dipole (E1) approximation. This excitation rate is sufficient to determine the detector efficiency, assuming an efficient transition from the excited state so that the success of the excitation process is detected through a click. This allows predicting in detail the mapping of a LG beam for various types of detectors (sub-section 3.2).

## 3.1. Excitation of an elementary quantum system by optical fields with nontrivial spatial structure : beyond the dipole approximation.

We need to derive the excitation rate of an elementary quantum system by an optical field whose space structure is arbitrary. Using usual minimal coupling Hamiltonian for atom-field interaction [15] and Fermi golden rule [16], the excitation rates can be expressed through spatial and temporal correlation functions of the field amplitudes and their gradients at the atom position $r_0$ - and needs to be multiplied by atomic matrix elements. In general, the spatial behavior of the correlation functions depends on the quantized field initial conditions and should be stated by special theoretical and experimental criteria (for details, in the case of a single electron system, see [17]). In the following, we only consider optical fields in a coherent initial state, allowing correlation functions to factorize, and we assume a steady-state regime for the monochromatic field. Moreover, we consider here an elementary quantum system ("atom") assumed to be motionless (translation and rotation as well), and the transitions operators (of various ranks) are considered as given. Indeed, for reasons already mentioned in [9], what we have in mind for the detector is a complex ion or molecule embedded in a macroscopic solid matrix, rather than a free hydrogen atom, as considered in many models [18, 20, 22]. Note that, as considered in detail in [10], it is legitimate to neglect those transfer to the center of mass when one addresses only the internal transitions, although there always exists an angular momentum transfer to the center of mass, in the same way as atomic recoil is never absolutely null.

Assuming the transition to be nearly resonant with the monochromatic exciting field, the transition probability is expressed by

$$R = \frac{|T|^2}{\hbar} \frac{\Gamma/2}{\Gamma^2/4} \qquad (7)$$

$$T = T_{E1} + T_{M1} + T_{E2} = \mathbf{d} \cdot \mathbf{E}(\mathbf{r},\omega_0) + \mathbf{m} \cdot \mathbf{B}(\mathbf{r},\omega_0) + Q_{jk} \nabla_j E_k(\mathbf{r},\omega_0)$$

where **d** and **m** and $Q_{j,k}$ are respectively the electric dipole vector, magnetic dipole vector, and matrix elements of the quadrupole operator (NB: in [9], eq.1 is incorrect and differs from present eq.(7), but the mistake affects the situation only for $\delta\omega \neq 0$)..

In (7), the subscripts ($j,k = x,y,z$) denote Cartesian coordinates and are to be summed over when repeated. The resonance in the transition appears through the denominator, where $\delta\omega$ and $\Gamma$ are respectively the detuning between the field frequency $\omega$ and the atomic transition frequency $\omega_0$, and the width of the transition. It is here assumed for simplicity that the relevant detunings are the same for the E1, M1 and E2 transitions. Note that if in most cases, a strong resonance occurs for only one type of transition, it is not unlikely that a resonant transition mixes-up two kinds of transition in a complex atomic system, a typical example being chirality-sensitive transitions, with E1 - M1 mixing.

To obtain specific predictions, including the relevant selection rules, we consider a detection scheme based upon a single atom or molecule with a fixed orientation in space, (*i.e.* the reasoning does not apply to a distribution of randomly oriented atoms). For such a detector with an arbitrary internal structure, the nonzero elements of matrix **d**, **m**, and tensor $Q_{ij}$ are respectively governed by 3, 3 and 5 independent components. These matrix elements can be parameterized with Cartesian co-ordinates ($x, y, z$), where $z$ is the quantization axis, in the following form:

$$\mathbf{d}^{(\pm 1)} = d^{(\pm 1)}(\pm 1, i, 0); \qquad \mathbf{d}^{(0)} = d^{(0)}(0, 0, \sqrt{2})$$
$$\mathbf{m}^{(\pm 1)} = m^{(\pm 1)}(\pm 1, i, 0); \qquad \mathbf{m}^{(0)} = m^{(0)}(0, 0, \sqrt{2})$$
(7)

$$\mathbf{Q}^{(0)} = Q^{(0)} \sqrt{\frac{2}{3}} \begin{Vmatrix} -1 & 0 & 0 \\ 0 & -1 & 0 \\ 0 & 0 & 2 \end{Vmatrix}; \quad \mathbf{Q}^{(\pm 1)} = Q^{(\pm 1)} \begin{Vmatrix} 0 & 0 & \mp \\ 0 & 0 & -i \\ \mp & & 0 \end{Vmatrix}; \quad \mathbf{Q}^{(\pm 2)} = Q^{(\pm 2)} \begin{Vmatrix} 1 & \pm i & 0 \\ \pm i & -1 & 0 \\ 0 & 0 & 0 \end{Vmatrix}$$

(7)

Through eq. (7) we have made explicit how the complicated fields and their spatial derivatives interact selectively with the corresponding multipole moments of atomic or molecular system [eqs. (7) and (7)].

### 3.2 Detection and mapping of a $|l|$ = 2 Laguerre-Gauss beam with elementary quantum systems

As already seen (section 2.3), the magnetic field and the gradient of the electric field of LG beam with $|l|$ = 2 are nonzero on the axis, in spite of a null electric field on-axis. This allows unusual effects to occur: in particular, the mapping of a LG field depends on the nature of the detector, with detectors of the magnetic field (M1 transition) and of the gradient of the electric field (E2 transition) yielding a non zero response on-axis. Our purpose here is to analyze quantitatively these effects. For sake of simplicity, we consider for our detector only an atom in a spherical ground state (S state), which can be excited through an E1, M1, or E2 transition to a Zeeman substate |L, M> (this limits M to M = 0, ±1, and L to $\Delta L = 0, \pm 1 (L = 0 \nleftrightarrow 0)$ for a dipolar transition E1 or M1, and to M = 0, ±1, ± 2 for a quadrupole E2 transition)

The analytical values of the transition amplitudes, calculated for the central spot (x = y = z = 0), are shown in tables 1-3. As already discussed in section 2.3, for an E1 transition, only the $|l| = 0, 1$ situation yields a signal at this location, owing to the fact that LG beams are hollow (for the electric field) when $|l| > 1$ (as discussed above, for $|l| = 1$, the longitudinal component is nonzero and depends on the waist size). Most of the values obtained in these tables depend upon the focusing and the radial complexity of the considered field, as characterized by $kw_0$ and the quantized number $p$. In a way analogous to the on-axis magnetic energy for a $|l| = 2$ (section 2.3), some of these transitions are effective only for a strong focusing, a high $p$ value being approximately equivalent to an increased focusing (*i.e.* increasing the gradients). It is for those coefficients requiring a strong focusing that the most unexpected results (i.e. allowing transitions forbidden with a plane wave excitation oriented along the z axis) are obtained. Conversely, some of these coefficients exist already for a plane wave, owing to a standard longitudinal gradient, as it is the case for the nearly forbidden M1 and E2 transitions. For E2 transitions (table 3), the sign of some coefficients can vary, corresponding to a situation in which the focusing opposes to the effect of the complicated radial structure. Actually, it is not a severe restriction to assume $kw_0 \geq 5$ (see fig.2), making the effect of focusing dominant in most practical cases, and negative the coefficients $C_{E2}^{\pm 1,0}$ and $C_{E2}^{0,\pm 1}$.

The mixture between spin and orbital momentum becomes apparent when considering circular polarizations (*i.e.* for $\alpha = 1/\sqrt{2}, \quad \beta = \pm i/\sqrt{2}$ ) when σ has the meaning of the longitudinal component of the spin of the photon. In this case, the matrix elements $\mathbf{m}^{(M)} \cdot \mathbf{B}^{(l)}$, $\mathbf{d}^{(M)} \cdot \mathbf{E}^{(l)}$, $\nabla \cdot \ddot{\mathcal{Q}} \cdot \mathbf{E}$ are nonzero only for M = *l*-1 for a σ = -1 polarization, and for M = *l*+1 for a σ = +1 polarization, showing that that z-component of angular momentum $j_z = \hbar$ of the LG beam has been transferred to the atom. This result appears in agreement with the independent approximate calculations of angular momentum of the beam as in eq. (7) and with the conservation of angular momentum. In the case of an E2 transitions, the total angular momentum of the atom increases from $0\hbar$ to $2\hbar$. Note that one could also predict transitions with ±$3\hbar$ increase for *l* = +2 and σ = +1 (or *l* = -2 and σ = -1), provided that one extends equation (19) to include higher order multipolar terms.

Tables 1-3 already demonstrate how the *relative* efficiency of various types of detectors differs with the focusing, and with the polarization. We discuss below the transversal spatial distribution of these excitation rates for multipole detectors (E1, M1, E2) located in the waist plane of a LG beam. For this purpose, figures 7-9 show these distributions for beams with *l* = +2 and bearing different polarizations. A typical situation of strong focusing $kw_0$ = 6 and highly structured beam (*p* = 6) has been chosen, to emphasize the radical differences between the various mappings.

In figure 7, corresponding to a circular polarization σ = - 1 leading to $j_z$ = 1, the various "images" of the beams differ notably with respect to the apparent size of the LG beam but still exhibits cylindrical symmetry. M1 and E2 transitions look comparable for a comparable sub-state transition, although some quantitative differences occur in the radial dependence. The "hollow region" looks smaller for M = 0 than for a M = -1 transition, whatever is the nature of the transition This can be understood qualitatively, at least for the case of E1 and M1 transitions by noting that, in figure 3, the central node for $E_x$, $E_y$ (and $B_x$, $B_y$), is much broader than the one for $E_z$ (and $B_z$) : indeed, the M= -1 transition is purely connected to the transverse field, while the M= 0 is the signature of a coupling to the longitudinal field. As expected from the presence of a magnetic field and gradient of electric field for $l = 2$ (table 1), one can get a bright spot at the center for the M1 and E2 detectors, but only for M=+1 (see table 1, see also fig. 3 of ref. [9]). For a M1 transition, this is easily understood by the fact that the B field on center has a circular polarization governed by the sign of $l$ (= +2), independently of the sign of σ (=-1).

The figure 8, corresponding to the opposite circular polarization $\sigma = +1$, yielding $j_z = +3$, exhibits some remarkable differences with figure 7, in spite of various analogies (including a cylindrical symmetry). Among the latter, we note again a close similarity between the equivalent M1 and E2 transitions, and an increased diameter of the hollow regions when comparing (right to left on the figure) M=+1, M= 0, and M=1. For the E1 transition, as a simple result of standard selection rules, the transition to M= -1 (instead of M= + 1) is now forbidden. A major difference is associated to the combined effect of polarization and topological charge, *i.e.* to the $j_z$ value: for all comparable transitions, the "size" of the detected image is much less focused for $\sigma = +1$ than for $\sigma = -1$. This can be understood (for the E1 and M1 cases) by comparing figs 3 and 4, noting that the nodal regions of the longitudinal fields ($E_z$ and $B_z$) are larger than when σ =- 1. A remarkable consequence is that, even for a standard detection relying on an E1 transition and non selective regarding the Zeeman substates (*i.e.* averaging the mapping for M = -1, M = 0, M = +1), the apparent size of the LG beam differs for $\sigma$ = +1 and for $\sigma$ = -1, when the transverse field parameters are similar. In the case of an E1 transition, these differences are to be attributed to the effect of the longitudinal electric field. For M1 and E2 transitions, these size effects are naturally connected to the magnetic field structure, and gradient of electric field Another difference, already predicted, is that for a M1 transition with l = +2, there is no magnetic field for this specific case of polarization (see e.g. section 2), and hence no signal at the center. Also, there is no signal at center for an E2 transition. At this point, a general argument of angular momentum conservation can be used to understand these null signals at center, and the difference with figure 7. For a beam bearing 3ℏ units of angular momentum, any internal transition (under the restriction of an E2, or E1, M1 transitions) must be accompanied by a transfer of angular momentum to the center of mass of the atom: on center, such a transfer becomes prohibited for symmetry reasons.

These differences between the σ = -1 and σ = +1 situations are responsible for is even more unusual mappings when the focused LG beam is "linearly polarized" (with respect to the transverse component of the electric field). Indeed, figure 9 reveals a breaking of the cylindrical symmetry, the preferential orthogonal axes (x,y) being defined relatively to the polarization (x) of the electric field. This could be predicted from figs 5-6, as the field for a linear polarization, resulting from the summing of circular polarizations, exhibits a two-fold symmetry axis along fixed axes. For M1 and E2 transitions, this cylindrical symmetry breaking appears for all the M = 0, ±1 components; the cylindrical symmetry is however recovered for M = ±2, because only one circular polarization, yielding intrinsically a mapping with cylindrical symmetry, is active : only the σ = -1 polarization contributes to the M = -2 transition (and σ = +1 for the M = +2 transition). Even in the elementary case of an E1 detector, a symmetry breaking occurs, but only for M = 0. This indicates that a standard E1 detector, averaging over the Zeeman sub-components, provides a "picture" of the beam in which the direction of the polarization is highly recognizable.

Finally, this symmetry breaking appears to be of significance for the chiral properties of LG fields. This point will be addressed in a separate work.

*4. Conclusion*

In a general analysis of a broad validity, not limited to the paraxial beam approximation, nor to the electric dipole (E1) approximation, we have provided new insights on the spatial structure of complicated electromagnetic fields, and on the way to detect the peculiar features associated to these structures that vary rapidly on the scale of an optical wavelength. Our results exemplify the idea, counterintuitive to numerous opticians, that light intensity is not just connected to the electric field amplitude, a result true only for smooth beams, more or less equivalent to the far-field limit. The most salient features were obtained here for an extreme focusing of the LG beams that makes comparable the longitudinal and transverse structures of the field. This is why our results extend far beyond the standard frame of LG beams, and are valid for any beam bearing a complicated structure. In particular, all quasi-axially symmetric fields with |*l*|= 2 should have non zero magnetic fields and gradient of electric fields at the axis, so that, strictly speaking, they are not "hollow beams". Our results, showing differences according to the type of detector used for the mapping, could help to characterize unusual solutions of Maxwell equations, such as knotted or point–like fields [19].

For sake of simplicity, we have considered here coherent fields, but the frame of our discussions can be easily extended to any quantum state of the exciting field. This should enable one to analyze high-order quantum optics correlations, notably between differing types of detectors located at spatially separated points. This may bring in particular new tools to benefit of the extended set of entanglement variables [6] offered by LG beams and their angular momentum. In particular, the prediction of a different mapping for the two circular polarizations could open new prospects when one considers the spatially separated detection of twin photons of opposed circular polarizations.

Our analysis was initially triggered by the need to understand better the long sought effects of the transfer of orbital momentum to atoms, notably leading to the transfer of several units of angular momentum (in a combination of spin and orbital momentum) in a single interaction between a photon and an elementary quantized particle (*i.e.* an atom), as it occurs when a LG beam irradiates a resonant atomic medium in a linear absorption experiment [10,20,21 ]. Our detailed analysis confirms that a transfer of several units of the longitudinal component of the orbital angular momentum can occur, provided that the appropriate multipolar transition is considered. This transfer, obeying selection rules [10], depends upon the position as well as the orientation of the quantum (atomic) detector. For some transitions (M1 and E2 for $|l| = 2$, higher order transitions for $|l| > 2$), we even predict that the maximal probability is obtained on the axis of a beam often considered to be hollow. This is reminiscent of the Jáuregui prediction of a maximal spontaneous emission of "Bessel photon" on-axis [22]. On-axis, the allowed transitions are simply governed by the $j_z$ value, when $|\sigma| = 1$ (all other cases, including the linear polarization situation $\sigma = 0$, being mixed cases). It is when the detector is off-axis, that the selection rules turn out to be more complex, owing to the direction (and projection axis) of the absorbed photon.

Although our results are related to elementary properties of propagating fields, the implementation of an experimental demonstration of our predictions may remain difficult, even if the considered focusing ($kw_0 = 6$) cannot be considered as unrealistic. Production of LG beams from a standard $TEM_{00}$ laser is now a standard operation, but only in the paraxial limit, intrinsically not suitable to produce strongly focused beams [23]. Rather, it appears that the strong focusing of a collimated LG beam produces a beam which is no longer a LG beam, and whose polarization pattern is complex [24]. However, the development of nano-optics technology should enable to fabricate special optics (compatible with the spiral nature of the wave front) that would produce the tailored focused beams [25], that we have described as LG beams above the paraxial approximation. In addition, for the tight focusing that we consider here, the atomic detector (atom, or ion, or molecule, ...) should be a fixed particle located on a moveable plate, as in confocal microscopy, rather than the atoms of a vapor (a trapped ion is also considered recently [26]).

Finally, the major results that we present are obtained at locations where the propagating beams are so focused that there are far from being transverse. Rather, their complex structures represent a kind of near-field image of a source of a complex distribution of remote moving charges. With this respect, our study is closely related to the near-field regime, and should enhance the interest for the fabrication of nano devices mimicking LG behaviors, notably devices whose field include an angular momentum.

**Acknowledgements**

VK is grateful to the Russian Foundation for Basic Research (grants 11-02-91065, 11-02-92002, 11-02-01272, 12-02-90014, 12-02-90417) for partial financial support of this work and University Paris 13 for hospitality. DB MD and JRRL thank the French Brazilian (CAPES-COFECUB #456/04and Ph740/12) cooperation support. Work performed in the frame of International Program of Scientific Cooperation ("P.I.C.S." No 5813) between C.N.R.S. and Russian Foundation for Basic Research.

**Table 1.** Values of excitation amplitudes $T_{E1}^{lM}/E_0$ in the waist plane on-axis (x=y=z=0) (beam with an arbitrary polarization). One has defined $C_{E1}^{\pm 1,0} = -\dfrac{2\sqrt{2}\sqrt{p+1}}{\sqrt{\pi}kw_0}$

|  | M=-1 | M=0 | M=1 |
|---|---|---|---|
| $l=-2$ | 0 | 0 | 0 |
| $l=-1$ | 0 | $C_{E1}^{-1,0}(\alpha+i\beta)d^{(0)}$ | 0 |
| $l=0$ | $\dfrac{-i\sqrt{2}(\alpha-i\beta)}{\sqrt{\pi}}d^{(-1)}$ | 0 | $\dfrac{i\sqrt{2}(\alpha+i\beta)}{\sqrt{\pi}}d^{(1)}$ |
| $l=1$ | 0 | $C_{E1}^{+1,0}(\alpha-i\beta)d^{(0)}$ | 0 |
| $l=2$ | 0 | 0 | 0 |

Table 2. Same as table 2 for $T_{M1}^{lM}$. One has defined $C_{M1}^{-2,-1} = C_{M1}^{2,1} = \dfrac{4\sqrt{2}\sqrt{(p+1)(p+2)}}{\sqrt{\pi}w_0^2 k^2}$ ;

|  | M=-1 | M=0 | M=1 |
|---|---|---|---|
| $l=-2$ | $C_{M1}^{-2,-1}(\alpha+i\beta)m^{(1)}$ | 0 | 0 |
| $l=-1$ | 0 | $-C_{M1}^{0,-1}(\alpha+i\beta)m^{(0)}$ | 0 |
| $l=0$ | $\dfrac{-\sqrt{2}}{\sqrt{\pi}}(\alpha-i\beta)m^{(-1)}$ | 0 | $\dfrac{-\sqrt{2}}{\sqrt{\pi}}(\alpha+i\beta)m^{(-1)}$ |
| $l=1$ | 0 | $C_{M1}^{0,1}(\alpha-i\beta)m^{(0)}$ | 0 |
| $l=2$ | 0 | 0 | $C_{M1}^{2,1}(\alpha-i\beta)m^{(1)}$ |

Table 3. same as table 2 for $T_{E2}^{lM} w_0 / E_0$. One has defined

$$C_{E2}^{\pm 1,0} = i \frac{2\sqrt{2(p+1)}}{\sqrt{3}\sqrt{\pi}} \frac{\left(8p+8-3(kw_0)^2\right)}{k^2 w_0^2} ; C_{E2}^{0,\pm 1} = \frac{\sqrt{2}\left(8p+4-(kw_0)^2\right)}{kw_0 \sqrt{\pi}}, C_{E2}^{-2,-1} = C_{E2}^{2,1} = \frac{4\sqrt{2(p+1)(p+2)}}{\sqrt{\pi} kw_0}$$

|        | M=-2 | M=-1 | M=0 | M=1 | M=2 |
|---|---|---|---|---|---|
| $l=-2$ | 0 | $-C_{E2}^{-2,-1}(\alpha+i\beta)Q^{(-1)}$ | 0 | 0 | 0 |
| $l=-1$ | $\dfrac{4i\sqrt{p+1}}{\sqrt{\pi}}(\alpha-i\beta)Q^{(-2)}$ | 0 | $C_{E2}^{-1,0}(\alpha+i\beta)Q^{(0)}$ | 0 | 0 |
| $l=0$ | 0 | $C_{E2}^{0,-1}(\alpha-i\beta)Q^{(-1)}$ | 0 | $-C_{E2}^{0,-1}(\alpha+i\beta)Q^{(1)}$ | 0 |
| $l=1$ | 0 | 0 | $C_{E2}^{1,0}(\alpha-i\beta)Q^{(0)}$ | 0 | $\dfrac{4i\sqrt{(p+1)}}{\sqrt{\pi}}(\alpha+i\beta)Q^{(2)}$ |
| $l=2$ | 0 | 0 | 0 | $C_{E2}^{2,1}(\alpha-i\beta)Q^{(1)}$ | 0 |

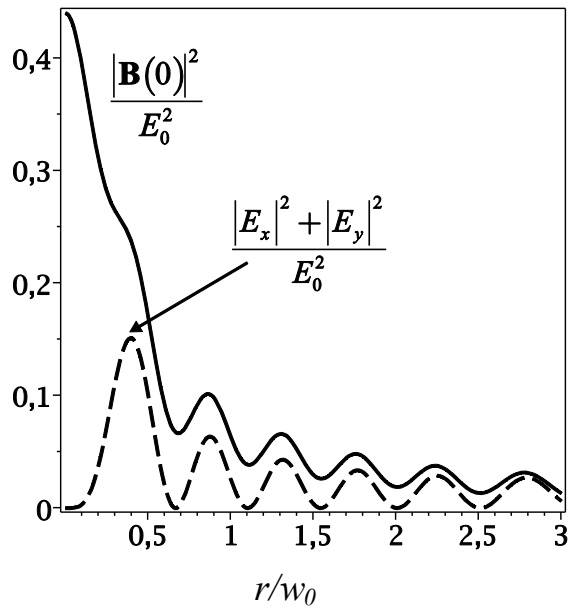

Fig.1. Radial dependence of the electric and magnetic energy density of a LG beam ($kw_0=6$, $p=6$, $l=+2$, $\alpha = 1/\sqrt{2}, \beta = +i/\sqrt{2}$)

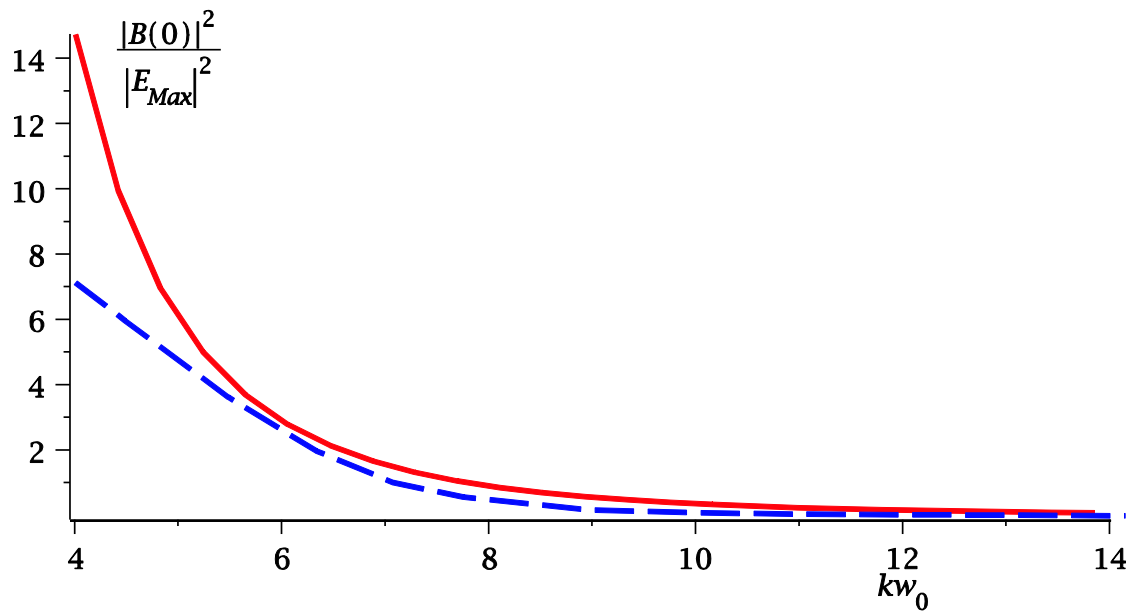

Fig.2. Ratio of the magnetic energy density at the center of a LG beam to the electric energy density of transverse field at its maximum, as function of beam waist $kw_0$ ($p=6$, $l=2$). The solid line corresponds to LG beam (see eq. 14), the dashed line is for a generalized Bessel beam (see eq.(4)-(5)). One can see from this figure that LG approximation is good enough until $kw_0=5$.

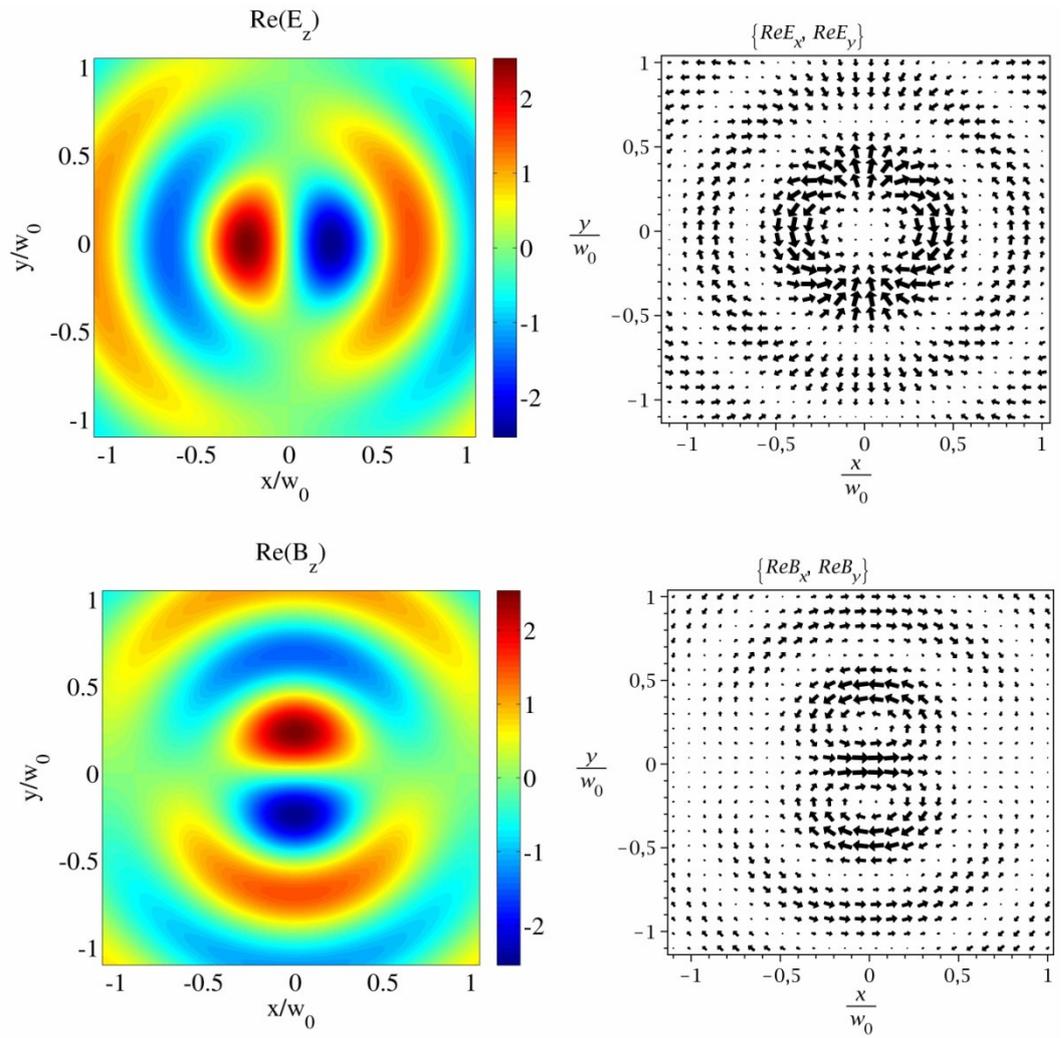

Fig.3. Spatial distributions of the real parts of the electric and magnetic fields in the waist of the LG beam ($kw_0=6$, $p=6$, $l=2$) with a circular polarization ($\sigma=-1$). The imaginary parts can be obtained by $90^0$ clockwise rotations.

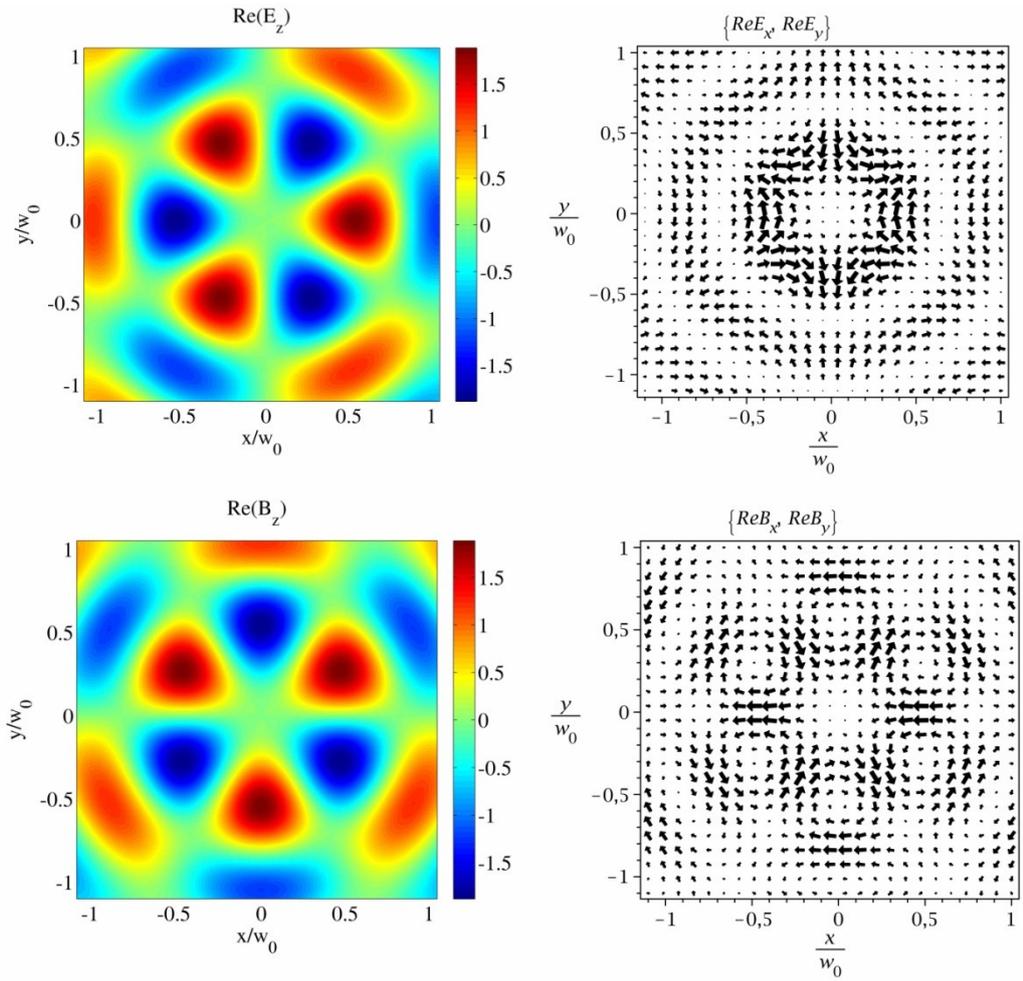

Fig.4. Spatial distributions of the real parts of the electric and magnetic fields in the waist plane of the LG beam ($kw_0=6$, $p=6$, $l=2$) with a circular polarization ($\sigma=+1$). The imaginary parts can be obtained by $90^0$ anticlockwise rotations.

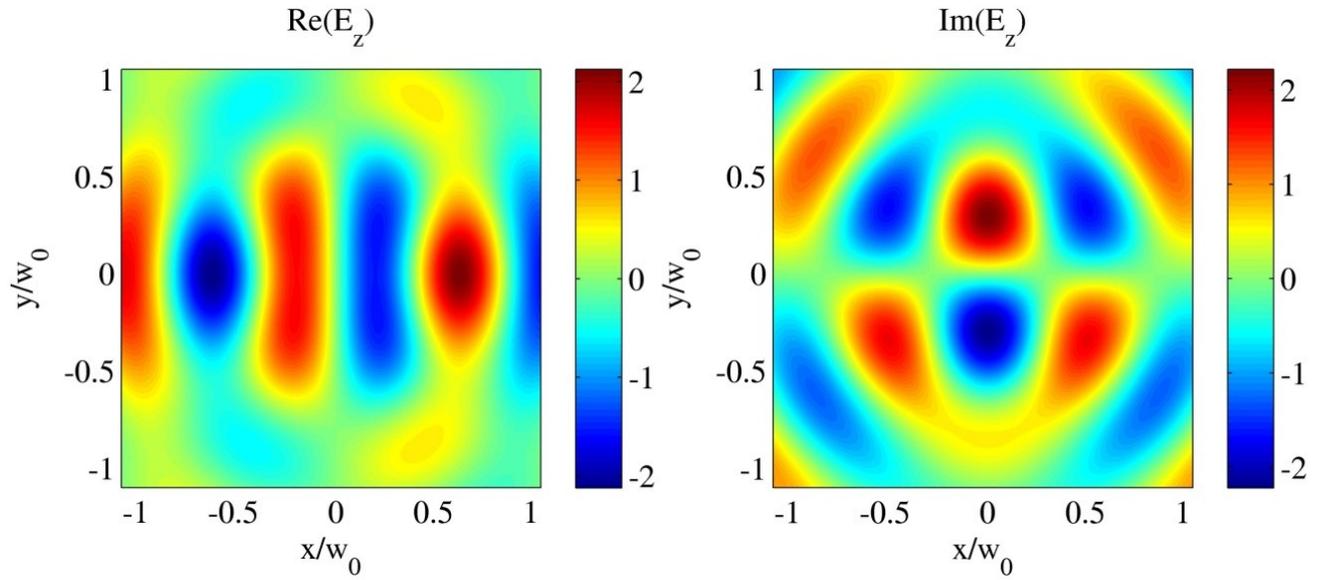

Fig.5. Spatial distributions of the longitudinal parts of the electric fields in the waist of the LG beam ($kw_0$=6, $p$=6, $l$=2) with a linear (*x*) polarization ($\alpha=1; \beta=0$). The distribution for magnetic fields can be obtained by $90^0$ clockwise rotation.

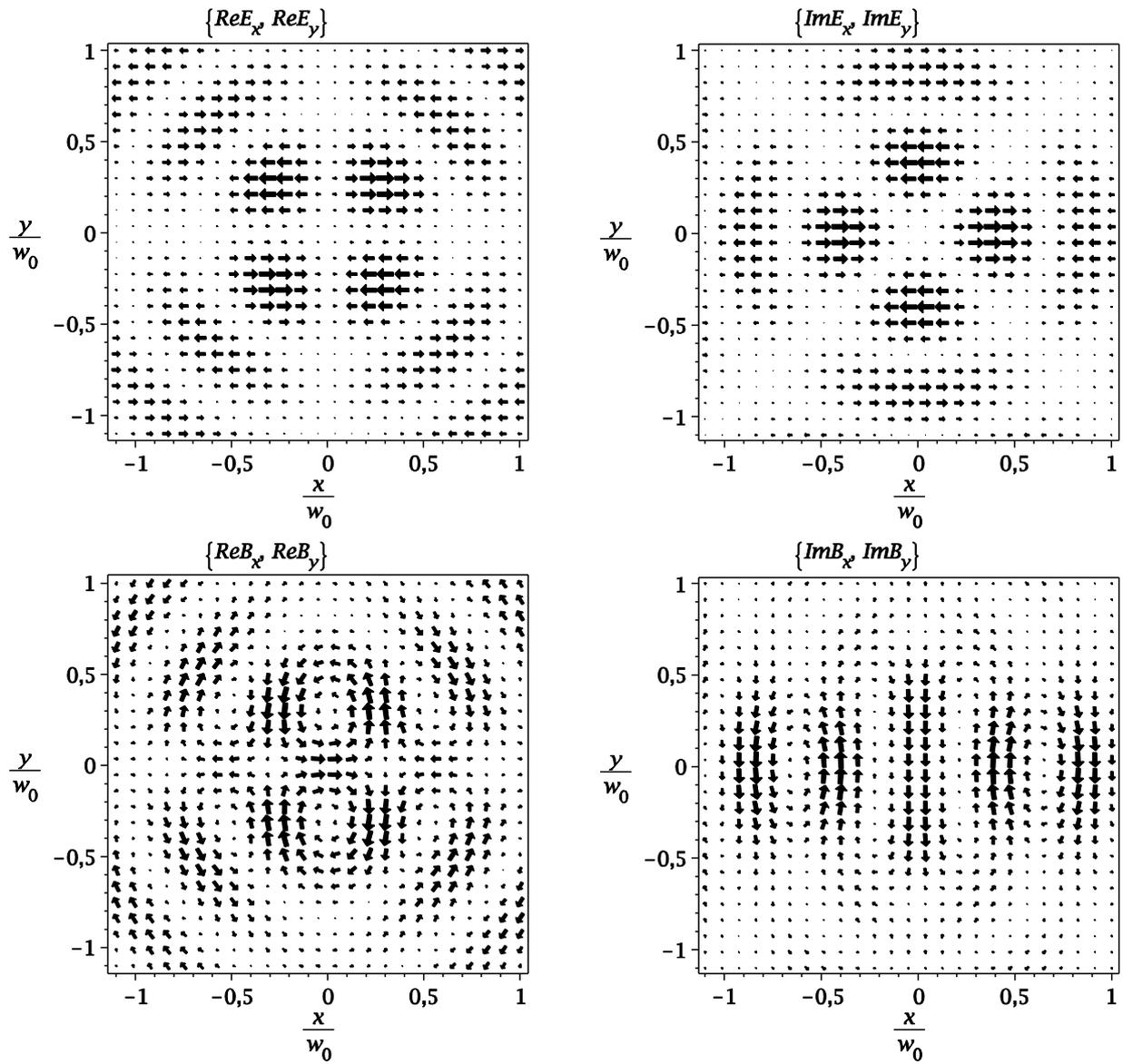

Fig.6. Spatial distributions of the transversal parts of the electric and magnetic fields in the waist of the LG beam ($kw_0=6$, $p=6$, $l=2$) with a linear (*x*) polarization ($\alpha = 1; \beta = 0$).

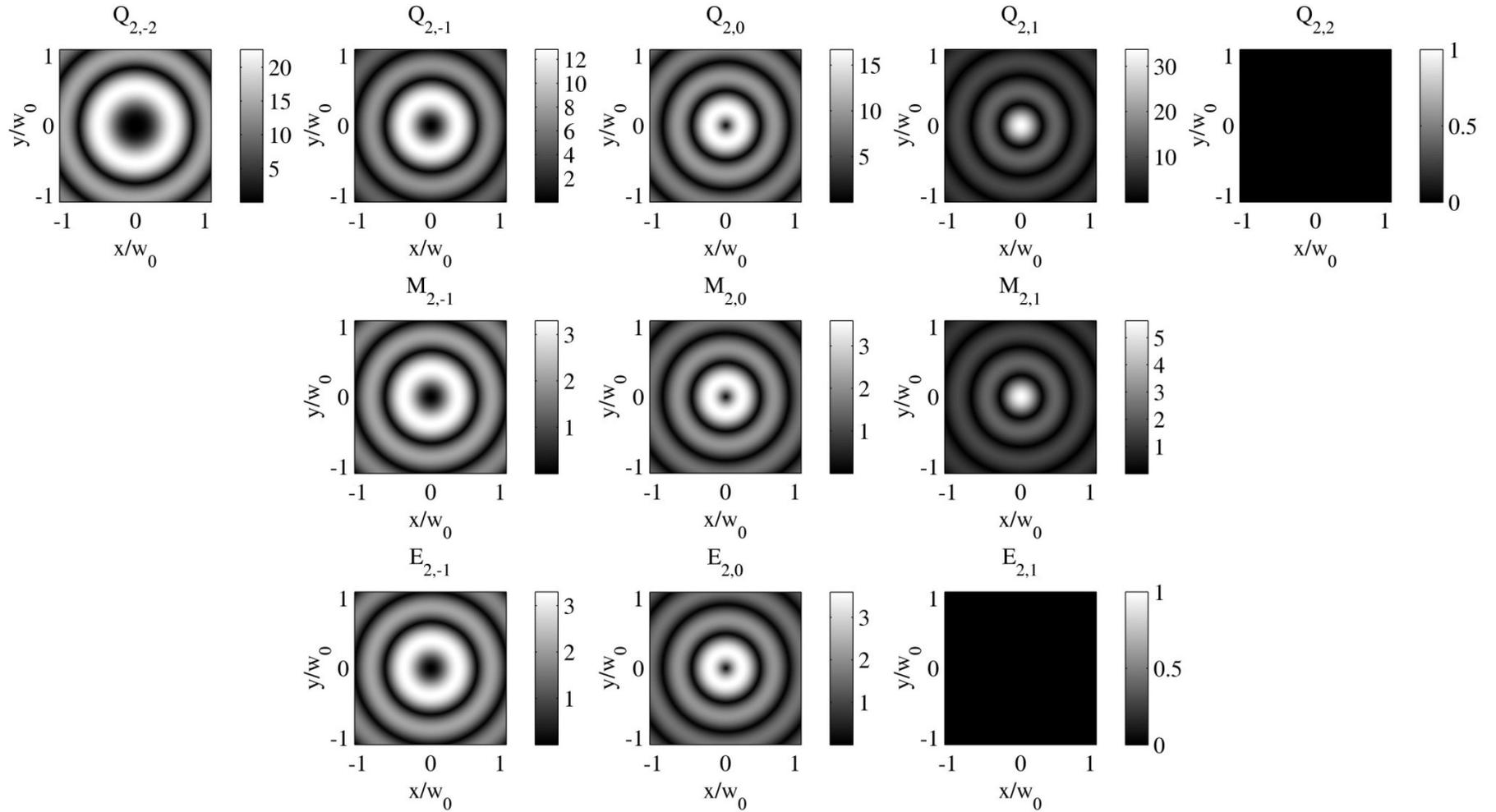

Fig. 7. Spatial distributions of the normalized excitation rates for an atom (detector) located in the waist plane of a circularly polarized ($\sigma$=-1) LG beam with $l$=2, $p$=6, $kw_0$=6. The upper row shows the electric quadrupole transitions rates, $\left(T_{E2}^{mM}/(E_0 Q^M)\right)^2$. The middle row shows the magnetic dipole transitions rates, $\left(T_{M1}^{mM}/(E_0 m^M)\right)^2$. The bottom row corresponds to the electric dipole transitions rates, $\left(T_{E1}^{mM}/(E_0 d^M)\right)^2$. Columns from left to right correspond to a transition to M = -2, -1, 0, +1, +2 for an E2-type detector, and to a transition to M = -1, 0, +1 for a M1 or E1 detector.

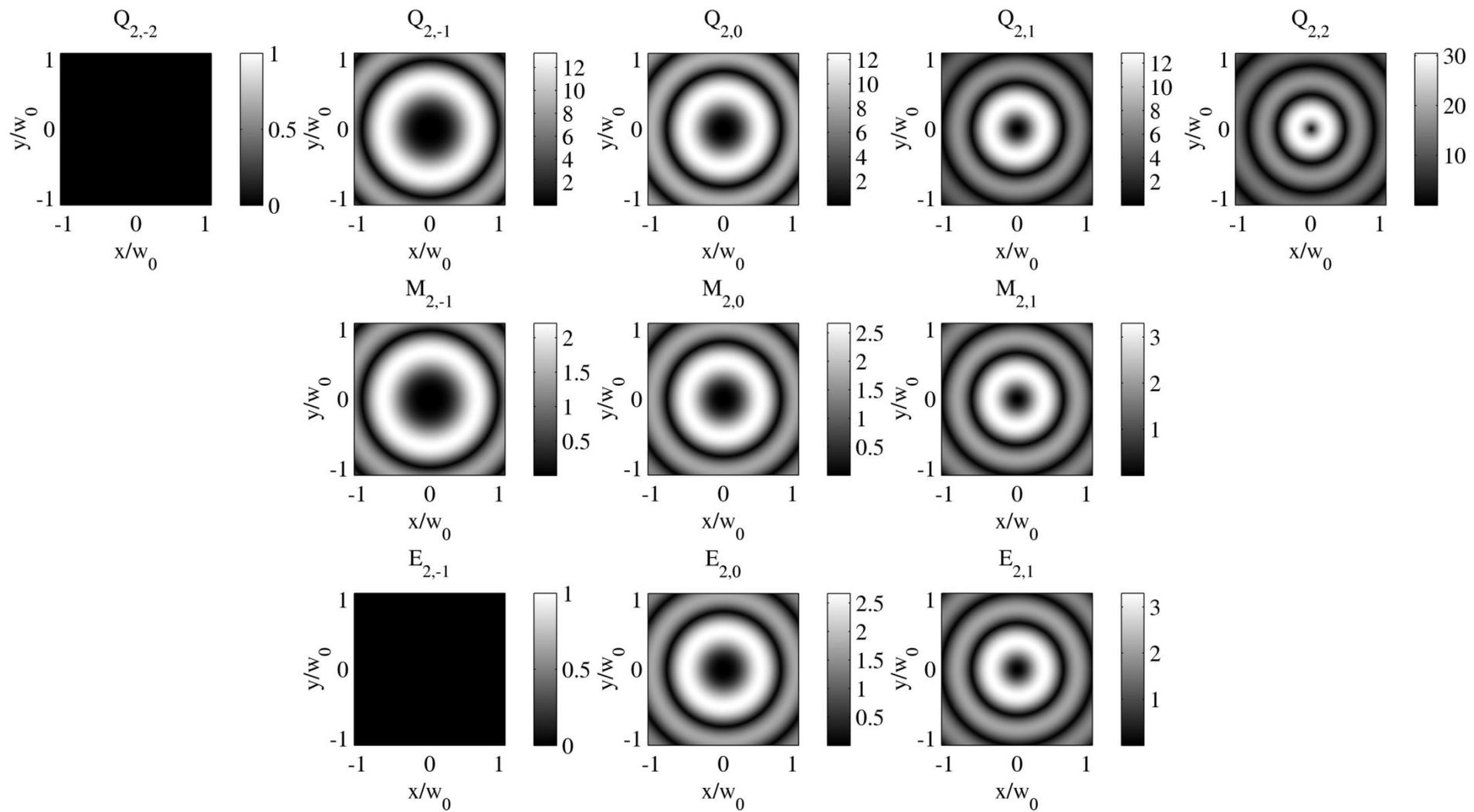

Fig. 8. Same as fig.7 for a circularly polarized ($\sigma=+1$) LG beam.

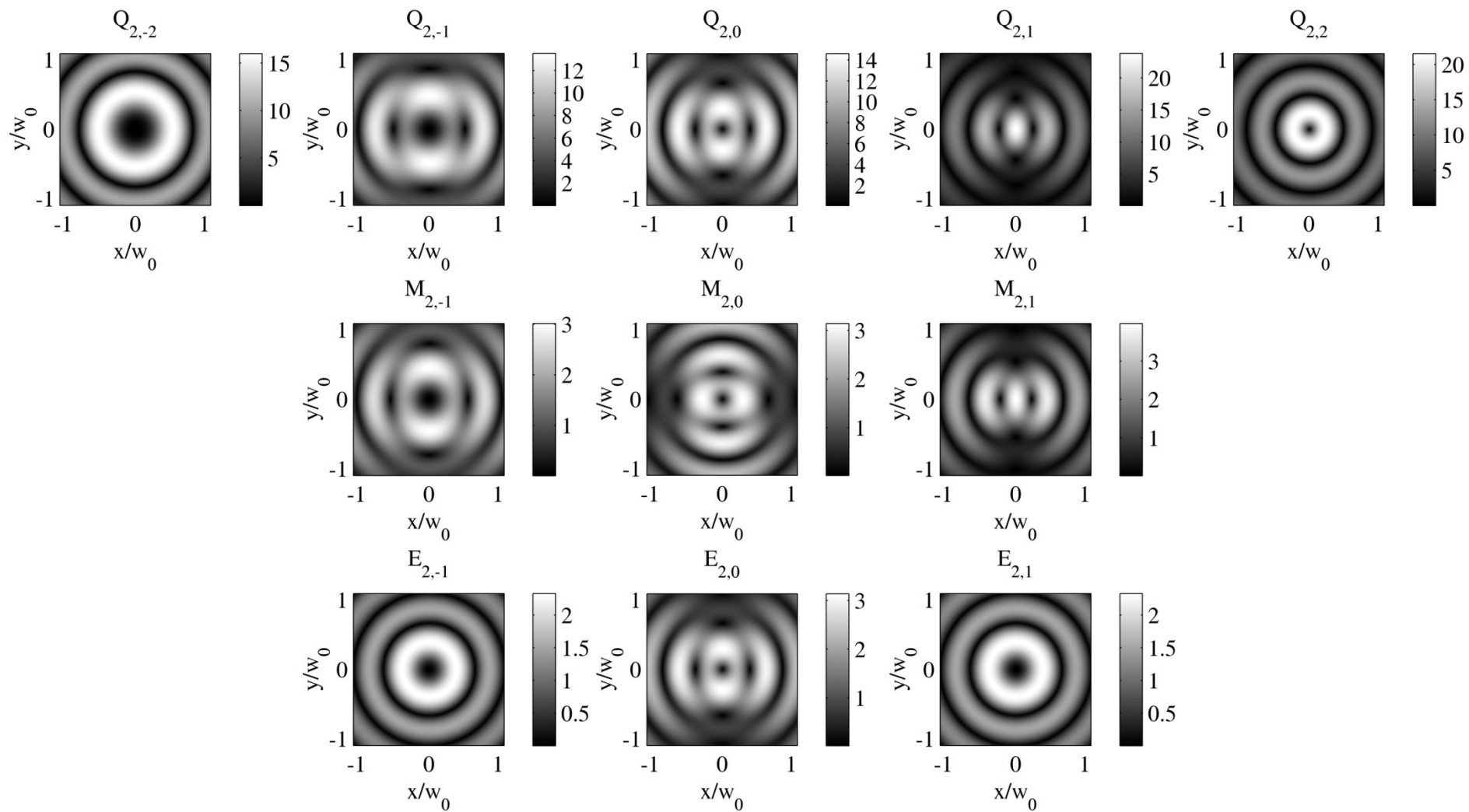

Fig. 9. Same as fig.7 for a linearly (*x*) polarized LG beam.

43